\documentclass[english,rsi,aip,superscriptaddress,reprint]{revtex4-2}
\usepackage[T1]{fontenc}
\usepackage[latin9]{inputenc}
\usepackage{color}
\usepackage{babel}
\usepackage{refstyle}
\usepackage{amsbsy}
\usepackage{graphicx}
\usepackage{microtype}
\usepackage[unicode=true,pdfusetitle,
 bookmarks=true,bookmarksnumbered=true,bookmarksopen=true,bookmarksopenlevel=1,
 breaklinks=true,pdfborder={0 0 0},pdfborderstyle={},backref=false,colorlinks=true]
 {hyperref}
\hypersetup{
 allcolors=blue}

\makeatletter

\AtBeginDocument{\providecommand\Figref[1]{\ref{Fig:#1}}}
\AtBeginDocument{\providecommand\Secref[1]{\ref{Sec:#1}}}
\RS@ifundefined{subsecref}
  {\newref{subsec}{name = \RSsectxt}}
  {}
\RS@ifundefined{thmref}
  {\def\RSthmtxt{theorem~}\newref{thm}{name = \RSthmtxt}}
  {}
\RS@ifundefined{lemref}
  {\def\RSlemtxt{lemma~}\newref{lem}{name = \RSlemtxt}}
  {}

\usepackage{newtxtext}
\usepackage[varg]{newtxmath}

\makeatother

\begin{document}
\title{Ray-tracing Analysis for Cross-polarization Scattering Diagnostic
on MAST-Upgrade Spherical Tokamak}
\author{R.~Hong}
\email{rohong@ucla.edu}

\affiliation{Physics and Astronomy Department, University of California, Los Angeles,
CA 90095, USA}
\author{T.~L.~Rhodes}
\affiliation{Physics and Astronomy Department, University of California, Los Angeles,
CA 90095, USA}
\author{G.~Wang}
\affiliation{Physics and Astronomy Department, University of California, Los Angeles,
CA 90095, USA}
\author{W.~A.~Peebles}
\affiliation{Physics and Astronomy Department, University of California, Los Angeles,
CA 90095, USA}
\begin{abstract}
A combined Doppler backscattering/cross-polarization scattering (DBS/CPS)
system is being deployed on MAST-U, for simultaneous measurements
of local density turbulence, turbulence flows, and magnetic turbulence.
In this design, CPS shares the probing beam with the DBS and uses
a separate parallel-viewing receiver system. In this study, we utilize
a modified GENRAY 3D ray-tracing code, to simulate the propagation
of the probing and scattered beams. The contributions of different
scattering locations along the entire beam trajectories are considered,
and the corresponding local $\tilde{\mathbf{B}}$ wavenumbers are
estimated using the wave-vector matching criterion. The wavenumber
ranges of the local $\tilde{\mathbf{B}}$ that are detectable to the
CPS system are explored for simulated L- and H-mode plasmas.
\end{abstract}
\maketitle

\section{Introduction\label{sec:Introduction}}

One significant feature of the spherical tokamaks (STs), such as MAST-U,\citep{harrison2019overview}
is the high ratio of the plasma kinetic energy to the magnetic energy
(i.e., high $\beta$). Although STs are more economically favorable
than conventional high-field tokamaks, the high $\beta$ may also
introduce more violent electromagnetic fluctuations and turbulent
transport. These magnetic fluctuations are predicted to be associated
with kinetic ballooning modes\citep{diallo2013observation} (KBMs)
and microtearing modes\citep{dickinson2012kinetic,smith2013characterization}
(MTMs), and can determine the edge plasma transport and thus pedestal
structure. Therefore, to characterize and predict the transport process
induced by these electromagnetic turbulence in high $\beta$ plasmas,
the local measurements of internal magnetic fluctuations are highly
desirable.

A novel diagnostic technique, combining Doppler backscattering (DBS)
and cross-polarization scattering (CPS), has been deployed on MAST-U.
This setup enables simultaneous measurements of local internal density
and magnetic fluctuations at the radial locations where KBMs and MTMs
have been predicted to be unstable in previous ST experiments. The
DBS diagnostic (a.k.a. Doppler reflectometry) measures intermediate-to-high
wavenumber density fluctuations and their propagation velocities.\citep{hirsch2001doppler,hillesheim2015intermediatekdensity,peebles2010anovel}
The CPS technique, on the other hand, measures internal magnetic fluctuations
based on the scattering of an incident microwave beam into the orthogonal
polarization by magnetic fluctuations.\citep{barada2016measurement,hillesheim2015intermediatekdensity,rhodes2014development,zou1995internal}
Specifically, the perturbed current induced in the plasma by the incident
electromagnetic wave, $\mathbf{E}_{i}$, can be written as\citep{lehner1989linearmode,wang2018firststep}
\[
\mathbf{J}=\frac{\tilde{n}}{n}\boldsymbol{\sigma}\cdot\mathbf{E}_{i}+\frac{\omega_{ce}}{\epsilon_{0}\omega_{pe}^{2}}\boldsymbol{\sigma}\cdot\left[\left(\boldsymbol{\sigma}\cdot\mathbf{E}_{i}\right)\times\frac{\tilde{\mathbf{B}}}{B}\right],
\]
where $\omega_{pe}$ is the plasma frequency, $\omega_{ce}$is the
electron gyro-frequency, $n$ is the local plasma density, $\mathbf{B}$
is the local magnetic field, $\epsilon_{0}$ is the vacuum permittivity,
and $\boldsymbol{\sigma}$ is the unperturbed conductivity tensor.
Here, the first term on the right-hand side represents the scattered
current of the same polarization (parallel to $\mathbf{E}_{i}$) by
density fluctuations; the second term corresponds to the CPS current
(orthogonal to $\mathbf{E}_{i}$) due to the magnetic fluctuations.
One may note that the unperturbed conductivity tensor is reduced to
$\sigma_{0}=ine^{2}/m_{e}\omega_{i}$ in isotropic plasmas. So, the
CPS wave power depends not only on local magnetic fluctuation but
also the local mean density.

\begin{figure}
\includegraphics[width=2.5in]{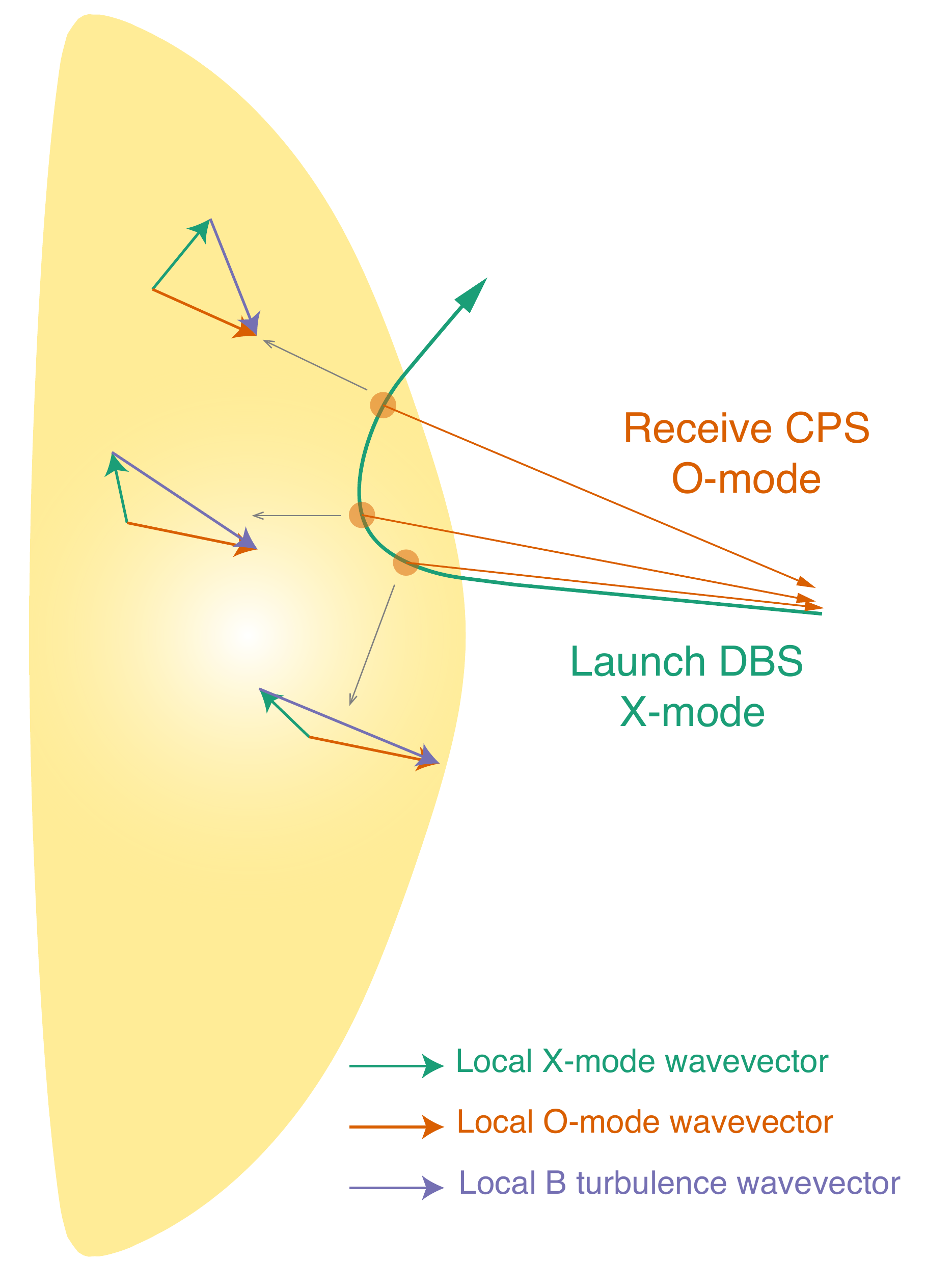}

\caption{\label{fig:diag-cps}Schematic of the cross-polarization scattering
processes for combined DBS/CPS system on MAST-U. The incident X-mode
mm-wave (green) launched by the DBS is scattered into O-mode CPS (orange)
at three different locations. The wave-vector matching conditions
for the local $\tilde{\mathbf{B}}$ (purple) are also shown.}

\end{figure}

The CPS diagnostic on the MAST-U shares the probing beam with the
DBS diagnostic, but uses a separate, parallel-viewing receiver antenna.
In principle, the CPS process occurs along the entire beam trajectory
once the wavevector and frequency matching conditions are satisfied.
The frequency matching condition is satisfied easily, since the frequency
of magnetic turbulence is about 3\textendash 4 orders of magnitude
less than that of the probing mm-wave, i.e., $\omega_{O}=\omega_{X}\pm\omega_{B}\approx\omega_{X}$,
where $\omega_{O}$ is the O-mode frequency, $\omega_{X}$ is the
X-mode frequency, and $\omega_{B}$ is the frequency of the matched
magnetic turbulence. The wavevector matching, however, is not trivial
for the CPS process. \Figref{diag-cps} demonstrates examples in which
the incident X-mode DBS wave was scattered into the O-mode CPS wave
at three different locations. To be received by the CPS system, the
wavevector of the scattered CPS wave ($\mathbf{k}_{O}$) should aim
towards the CPS antenna. At a given scattering center, if the wavevector
of the incident DBS wave ($\mathbf{k}_{X}$) is known, then the wavevector
of the local $\mathbf{\tilde{B}}$ turbulence ($\mathbf{k}_{B}$)
can be obtained via the wavevector matching condition, i.e., $\mathbf{k}_{B}=\mathbf{k}_{O}-\mathbf{k}_{X}$.
The matched wavenumber for the local $\mathbf{\tilde{B}}$ turbulence
varies substantially as the scattering location is changed (\Figref{diag-cps}).
As a result, the interpretation of the CPS measurements can be challenging.
A thorough understanding of the trajectories of incident and scattered
microwave beams is required to evaluate the probed wavenumber and
the location of the scattering centers. 

In this paper, we present the calculation of the wavenumber values
at multiple scattering locations by using modified GENRAY,\citep{smirnov1994ageneral}
a 3D ray-tracing code, for the combined DBS/CPS system on the MAST-U.
The wavenumbers for matched local magnetic turbulence are explored
in both L-mode and H-mode plasmas. Some effects and challenges that
may affect the CPS signals are also briefly discussed.

\section{GENRAY ray-tracing Analysis for combined DBS and CPS diagnostics}

\begin{figure}
\includegraphics[width=0.9\linewidth]{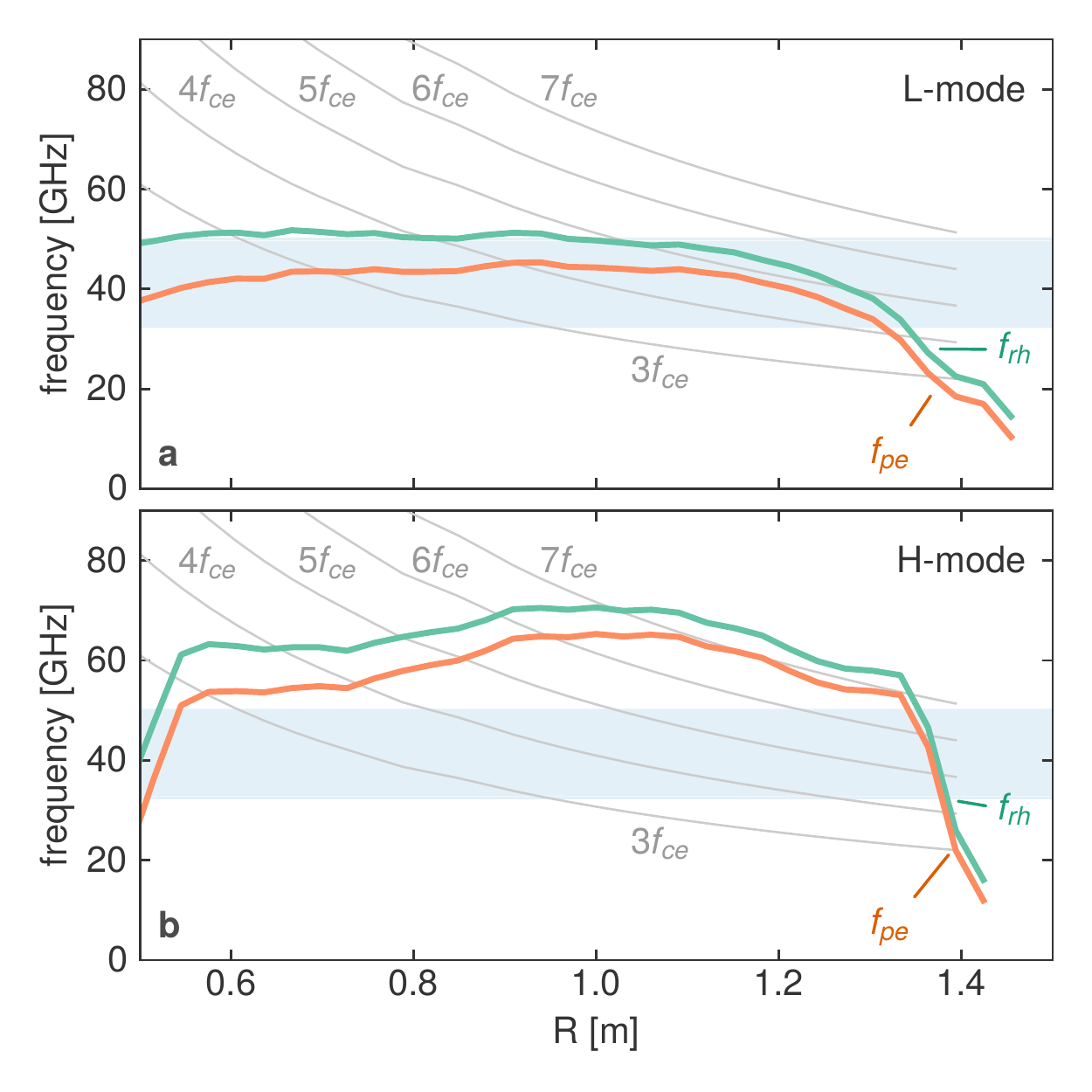}

\caption{\label{fig:cutoff-prof}Cut-off frequencies for the X-mode (green
$f_{\mathrm{rh}}$) and O-mode (orange $f_{\mathrm{pe}}$) plotted
against major radius of the device for L-mode (a) and H-mode (b) plasmas.
The shaded area indicates the frequency range covered by the DBS probing
beams.}
\end{figure}

The typical cutoff and resonance frequencies that may be present in
MAST-U\citep{harrison2019overview} are shown in \Figref{cutoff-prof}.
The right-hand frequency ($f_{\mathrm{rh}}$) and the plasma frequency
($f_{\mathrm{pe}}$) correspond to the cutoff frequency for the X-mode
wave and O-mode wave, respectively. The shaded area indicates the
frequency range of the DBS launch waves. In an L-mode plasma, the
probing beam covers the mid-radius region of the plasma, and the cutoff
radii for the X-mode ($f_{\mathrm{rh}}$) and O-mode ($f_{\mathrm{pe}}$)
beams are well separated (\Figref{cutoff-prof}(a)). The O-mode launch
will lead to a measurement of inner region in the L-mode plasma. In
the H-mode plasma, the probing beam can only cover the steep gradient
region at the pedestal, and the cut-off radii of the X-mode and O-mode
are close (\Figref{cutoff-prof}(b)). One may note that, if an O-mode
wave is launched to the plasma from the low-field side, the scattered
X-mode wave can only be receivable when the radius of the cross-polarization
scattering is greater than the right-hand cut-off radius. This indicates
that the effective measurement region of the CPS diagnostic is the
outer region of the plasma with its inner bound at the cutoff layer
of the X-mode wave.

In this study, the wavenumbers of the local magnetic turbulence measured
by CPS are analyzed using ray-tracing calculations based on a 3D ray
tracing code, GENRAY.\citep{smirnov1994ageneral} An example of procedure
is described as follows. First, an X-mode DBS probing beam ($f=45$
GHz) is launched into the plasma at a certain angle with respect to
the horizontal plane, and here four degrees of vertical angles are
chosen, i.e., $\theta\in[6^{\circ},\,8^{\circ},\,10^{\circ},\,12^{\circ}]$.
Since the scattering into O-mode can occur at any position along the
entire DBS ray, we then choose 7 scattering centers that are uniformly
distributed along each probing ray. In the next step, the O-mode CPS
ray is launched towards the CPS receiver at each of the scattering
centers. On the MAST-U the CPS receiver is at the same location as
the DBS launching system in the vessel ($R=2.28$ m and $Z=0$ m).
The ray-tracing calculations are iterated to ensure that these O-mode
rays can reach the CPS receiver.\citep{wang2018firststep} In our
calculations, we choose 2 mm as the minimum tolerable offset between
the CPS receiver and the end-points of each rays. The scattered wavevector
of the local magnetic turbulence at each $X\rightarrow O$ CPS locations
can be calculated using the wavevector matching conditions shown earlier,
i.e., $\mathbf{k}_{B}=(\mathbf{N}_{O}-\mathbf{N}_{X})k_{0}$. Here,
the local O- and X-mode wavevector $\mathbf{k}_{O}$ and $\mathbf{k}_{X}$
are written as $\mathbf{k}_{j}=\mathbf{N}_{j}k_{0}$, where $\mathbf{N}_{j}$
is the refractive index for O- or X-mode wave given by GENRAY calculations
and $k_{0}=2\pi f/c$ is the vacuum wavenumber. In following subsections,
we will show the results of the ray-tracing analysis in both L-mode
and H-mode plasmas.

\subsection{Ray-tracing results in L-mode plasmas}

\begin{figure}
\includegraphics[width=2.2in]{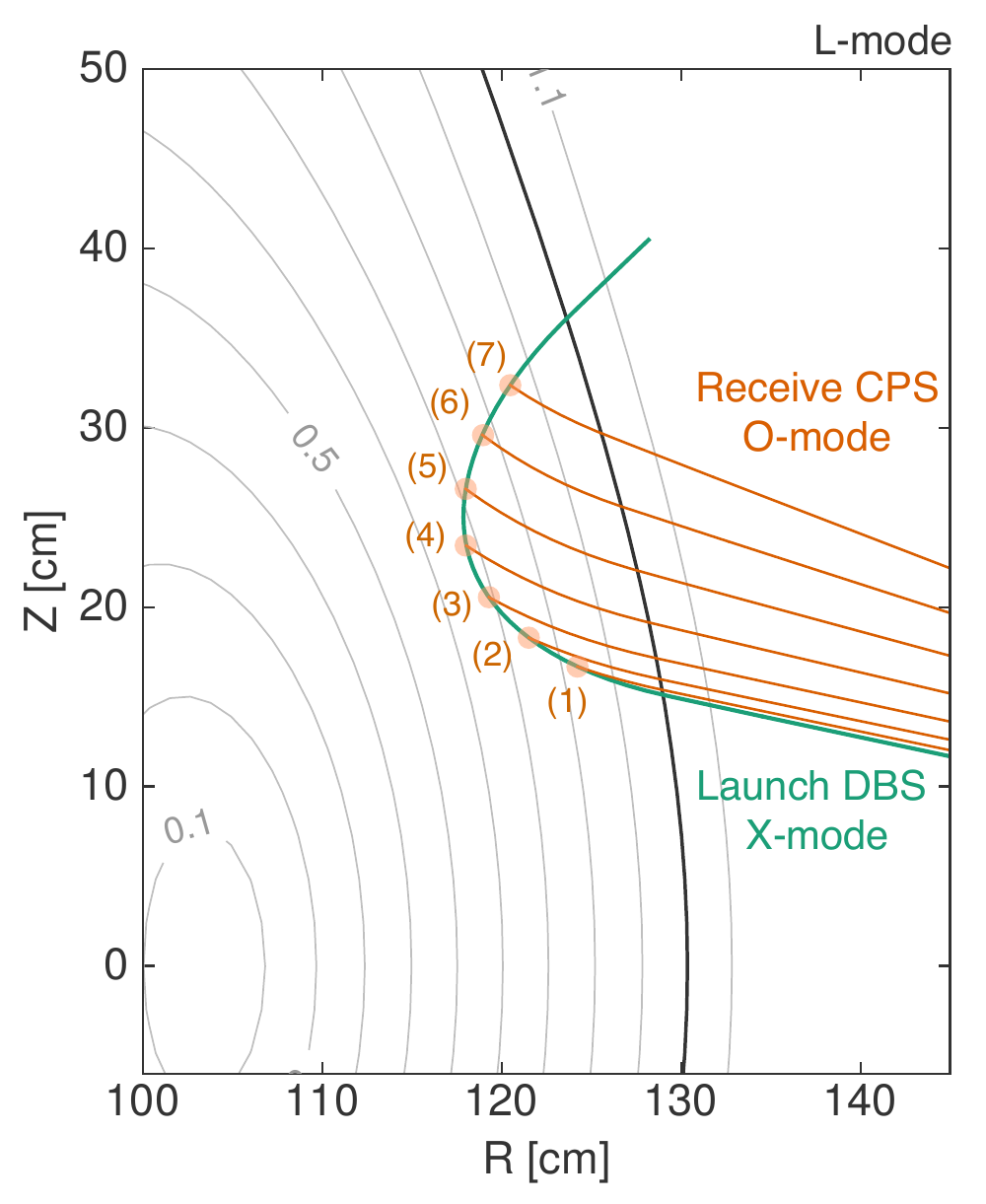}

\caption{\label{fig:Lmode}Ray trajectories in the R-Z plane of an L-mode plasma
in MAST-U, with the vertical launch angle $\theta=12^{\circ}$. 7
points uniformly distributed along the DBS probing beam are chosen
to analyze the wavenumber of the cross-polarization scattering waves
that can be received by the CPS antenna. The separatrix is highlighted
by the black curve.}
\end{figure}

\Figref{Lmode} shows the ray paths of the X-mode DBS probing beam
with the vertical launch angle $\theta=12^{\circ}$, as well as the
O-mode CPS rays in an L-mode plasma. The green curve represents the
trajectory of the incident X-mode DBS ray. The 7 chosen scattering
centers are uniformly distributed along the probing ray (orange circles
in \Figref{Lmode}), and all these points are inside the separatrix.
The O-mode CPS rays that can potentially be collected by the CPS receiver
are shown as orange curves in \Figref{Lmode}. The O-mode rays are
less refracted than the X-mode ray. Thus, the O-mode rays shown in
\Figref{Lmode} do not follow the X-mode ray back to the receiver. 

\begin{figure}
\includegraphics[width=0.95\linewidth]{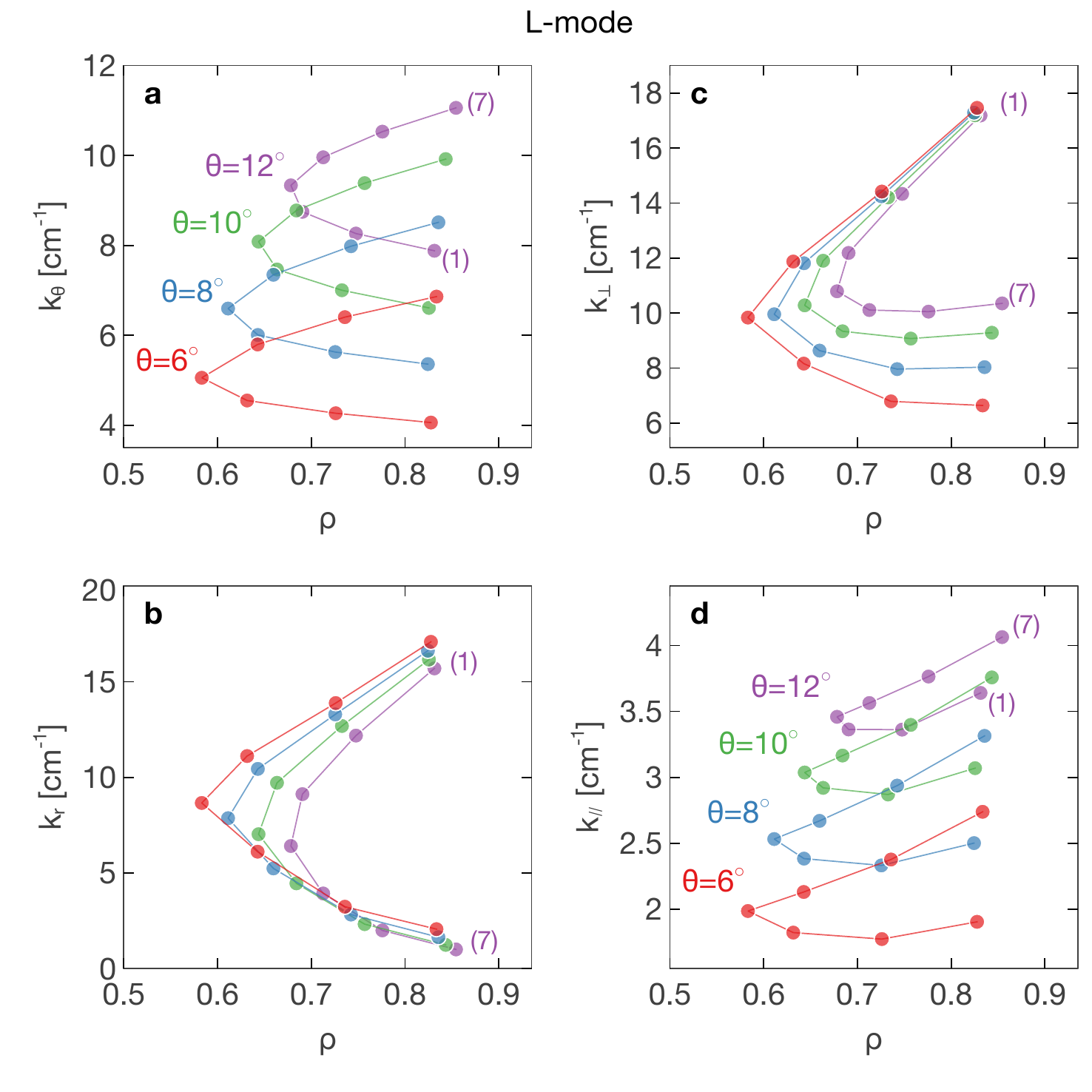}

\caption{\label{fig:cps_k_Lmode}Wavenumbers of magnetic turbulence in an L-mode
plasma in the (a) poloidal, (b) radial, (c) perpendicular, and (d)
parallel direction are shown, respectively. Each sub-figure shows
ray-tracing results for 4 different vertical launch angles of the
DBS ray, i.e., $\theta\in[6^{\circ},\,8^{\circ},\,10^{\circ},\,12^{\circ}]$.
7 scattering centers uniformly distributed along each DBS ray (corresponding
to points in \Figref{Lmode}) are chosen for the CPS ray-tracing calculation.}

\end{figure}

\Figref{cps_k_Lmode} shows the calculated wavenumbers of the local
magnetic turbulence in the L-mode plasma corresponding to four different
vertical launch angles of the probing ray. The 7 scattering centers
also corresponds to those displayed in \Figref{Lmode}. As the chosen
scattering center moves along the X-mode probing ray, e.g., from No.~1
to 7, the poloidal wavenumber increases (\Figref{cps_k_Lmode}(a))
and the radial wavenumber decreases (\Figref{cps_k_Lmode}(b)). In
other words, the matched wavevector of the local $\tilde{B}$ turbulence
is oriented more poloidally ($k_{B,\theta}\gg k_{B,r}$) at the later
part of the X-mode probing ray, and more radially ($k_{B,r}\gg k_{B,\theta}$)
if the CPS occurs at the early part of the probing ray. Here, both
of the radial and poloidal wavevectors lie in the perpendicular plane:
the radial wavevector is normal to the flux surface; the poloidal
wavevector is tangential to the magnetic flux surface and orthogonal
to both radial and parallel wavevectors, i.e., $\hat{k}_{\theta}=\hat{k}_{r}\times\hat{k}_{\parallel}$.
The perpendicular wavenumbers $k_{B,\perp}=\sqrt{k_{B,\theta}^{2}+k_{B,r}^{2}}$
are also greater when the CPS location at the later part of the incident
DBS ray (\Figref{cps_k_Lmode}(c)). The matched parallel wavenumber
does not change monotonically but the variation is relatively small
for different locations of the scattering centers (\Figref{cps_k_Lmode}(d)).

\subsection{Ray-tracing results in H-mode plasmas}

\begin{figure}
\includegraphics[width=2.2in]{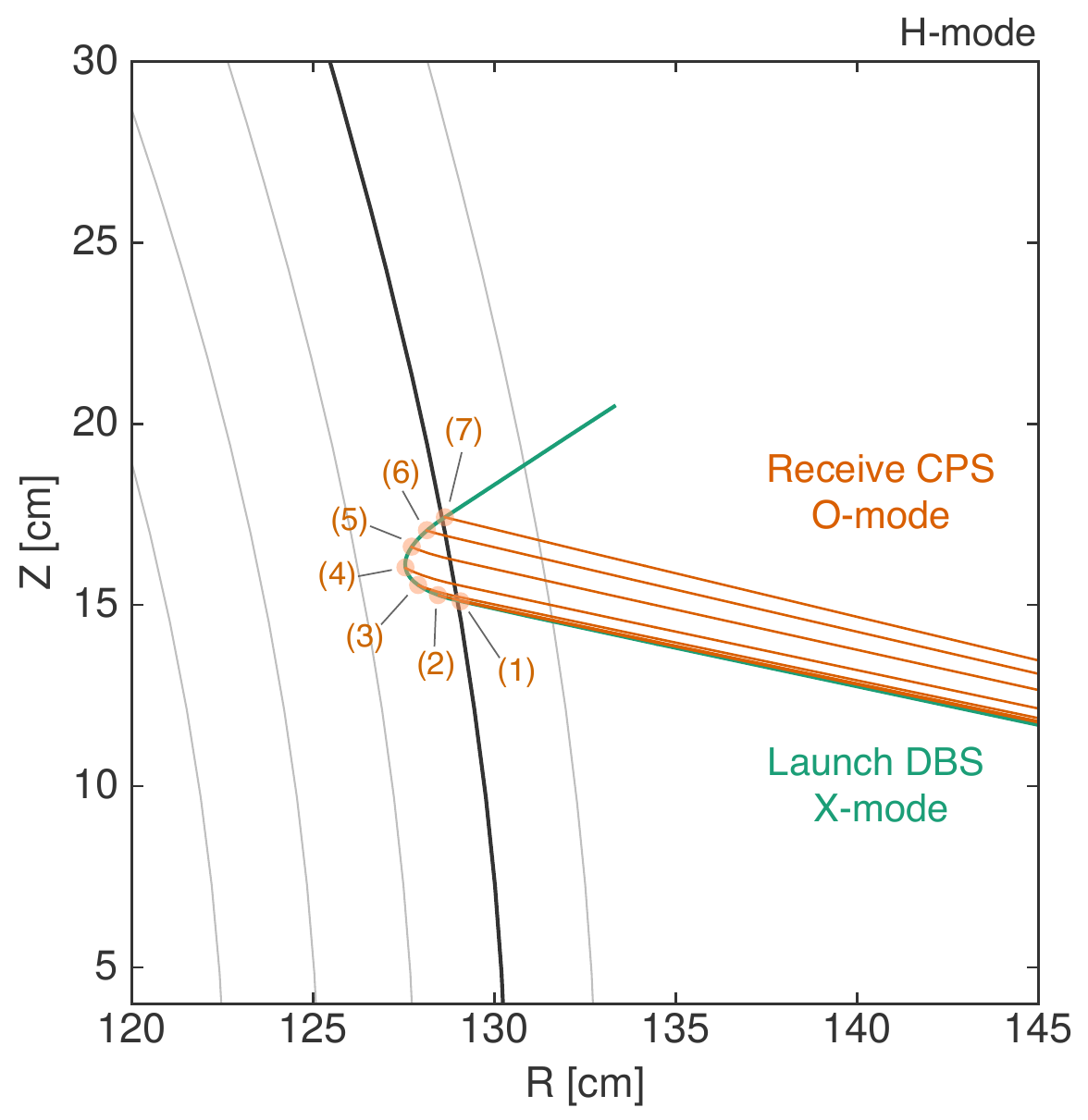}

\caption{\label{fig:Hmode}Ray trajectories in the R-Z plane of an H-mode plasma
in MAST-U, with vertical launch angle $\theta=12^{\circ}$. 7 points
along the DBS probing beam are chosen to analyze the wavenumber of
the cross-polarization scattering waves that can be received by the
CPS antenna. The separatrix is highlighted by the black curve.}
\end{figure}

\Figref{Hmode} shows the propagation of the X-mode DBS probing beam
with the vertical launch angle $\theta=12^{\circ}$, and the resultant
O-mode CPS rays in an H-mode plasma. Similar to L-mode case, the green
curve represents the trajectory of the incident X-mode DBS ray, and
the orange curves represent the detectable O-mode rays launched from
7 chosen $X\rightarrow O$ scattering centers. The scattering centers
are close to the separatrix due to the high density. One may also
note that the cutoff layer of the O-mode ray is close to that of the
X-mode ray in H-mode plasmas (\Figref{cutoff-prof}(b)). Some O-mode
CPS rays are almost overlapped with part of the X-mode probing ray
(No.~1\textendash 3 in \Figref{Hmode}), i.e., these CPS rays likely
follow the backscattering path of the X-mode ray. However, the CPS
O-mode ray emitted from the turning point of the incident X-mode ray
(No.~4 in \Figref{Hmode}) does not overlap with the backscattering
ray, although the O-mode cutoff is close to the X-mode cutoff in the
H-mode case (\Figref{cutoff-prof}(b)).

\begin{figure}
\includegraphics[width=0.95\linewidth]{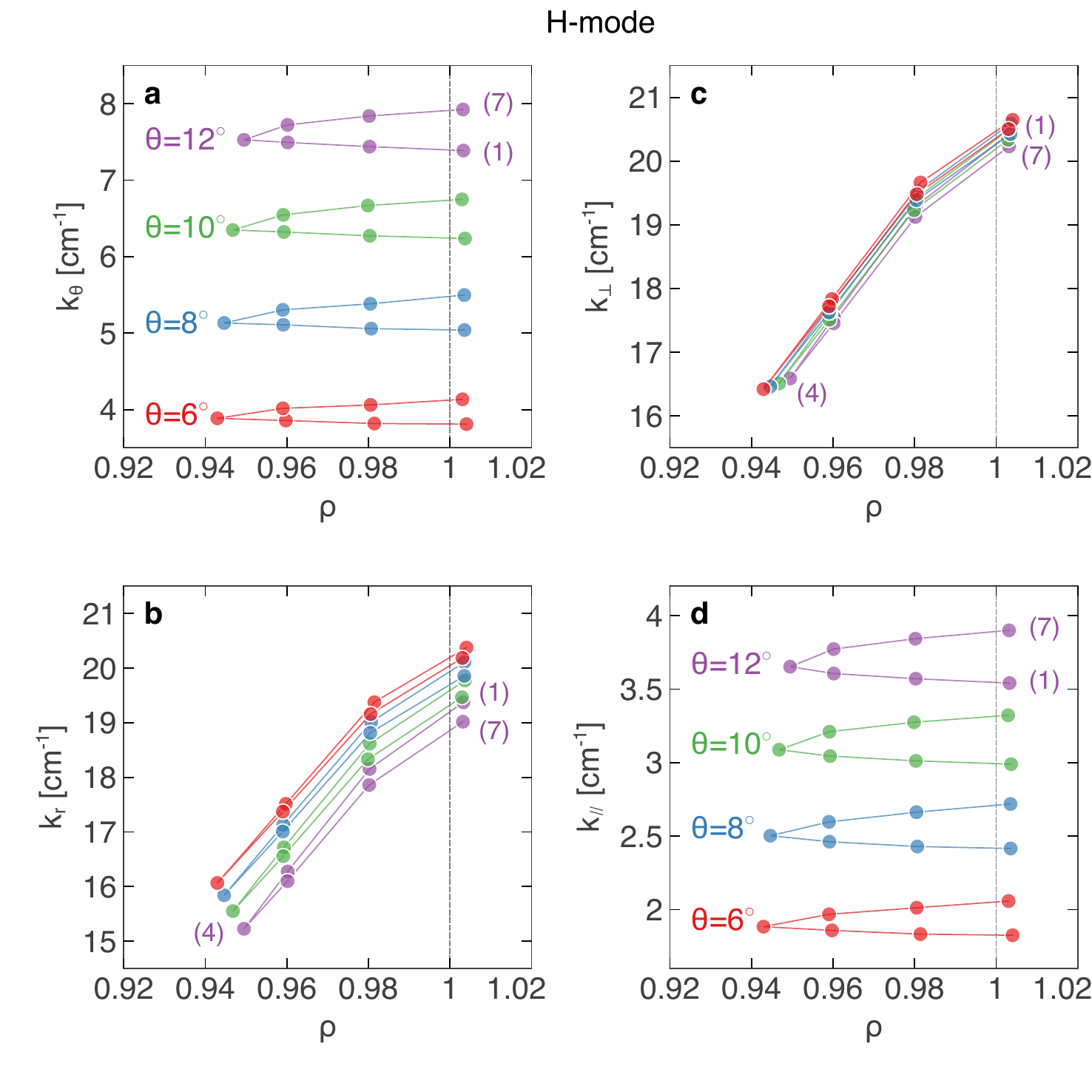}

\caption{\label{fig:cps_k_Hmode}Wavenumbers of magnetic turbulence in an H-mode
plasma in the (a) poloidal, (b) radial, (c) perpendicular, and (d)
parallel direction are shown, respectively. Each sub-figure shows
ray-tracing results for 4 different vertical launch angles of the
DBS ray, i.e., $\theta\in[6^{\circ},\,8^{\circ},\,10^{\circ},\,12^{\circ}]$.
7 scattering centers uniformly distributed along each DBS ray ((corresponding
to points in \Figref{Hmode})) are chosen for the CPS ray-tracing
calculation.}
\end{figure}

\Figref{cps_k_Hmode} shows the calculated wavenumbers of the local
magnetic turbulence in the H-mode plasma corresponding to four different
vertical launch angles of the probing ray. The 7 scattering centers
also corresponds to those displayed in \Figref{Hmode}. As the chosen
scattering center moves along the X-mode probing ray from No.~1 to
7, the poloidal wavenumber $k_{B,\theta}$ increases (\Figref{cps_k_Hmode}(a)),
but the radial wavenumber $k_{B,r}$ does not show a monotonic trend
(\Figref{cps_k_Hmode}(b)). Also, the $k_{B,r}$ is greater than $k_{B,\theta}$
by a factor of 2\textendash 3, and therefore the perpendicular wavenumbers
$k_{B,\perp}=\sqrt{k_{B,\theta}^{2}+k_{B,r}^{2}}$ is dominated by
the radial wavenumber $k_{B,r}$ (\Figref{cps_k_Hmode}(c)). The matched
parallel wavenumber does not change monotonically but the variation
is relatively small for different locations of the scattering centers
(\Figref{cps_k_Hmode}(d)).

\section{Some considerations on CPS signals\label{sec:Some-considerations}}

The signals received by the CPS system are determined by several factors.
Besides the local magnetic fluctuation and the mean density as discussed
in \Secref{Introduction}, the CPS signals are also affected by the
wavenumbers detected at the different scattering centers, the receiver
antenna pattern and orientation, the polarization misalignment, etc.
In this section, we briefly discuss these effects on the CPS signals
and their implications for our measurements.

The CPS signal level depends on the matched wavenumber of the local
magnetic turbulence. For example, the nonlinear gyrokinetic simulations\citep{guttenfelder2012simulation}
show that the magnetic fluctuations\textquoteright{} power spectrum
of the MTM turbulence is peaked at certain poloidal wavenumber range
($k_{\theta}\rho_{s}\sim0.2$) but obeys the power law for the radial
wavenumber spectrum ( $S_{B}(k_{r})\propto k_{r}^{-\alpha}$ where
$\alpha\approx3-4$). Hence, the strongest MTM contributions to the
CPS signals locate at $k_{B,\theta}\rho_{s}\approx0.2$ and $k_{B,r}\rho_{s}\approx0$.
Moreover, the parallel wavenumber of magnetic turbulence is much smaller
than its perpendicular counterpart, i.e., $k_{\parallel}\ll k_{\perp}$.
If the wavevectors of the launched and received microwave are \emph{not}
well aligned to match $k_{\parallel}\to0$, then the scattered power
will drop substantially. Specifically, for a Gaussian beam, this wavenumber
mismatch leads to a decrease in the scattered power by a factor of
$\exp\left[-\left(\delta k_{\parallel}\,a\right)^{2}/2\right]$, where
$\delta k_{\parallel}\approx\left(\mathbf{k}_{O}-\mathbf{k}_{X}\right)\cdot\mathbf{B}/\left|\mathbf{B}\right|$
is the mismatch of the parallel wavenumber and $a$ is the beam radius.\citep{slusher1980studyof}
Since the combined DBS/CPS diagnostic on MAST-U employs a 2D steering
optical system, the mismatch of the parallel wavenumber can be largely
reduced by using the toroidal steering.

The CPS signal level is affected by the receiver antenna pattern and
orientation as well. As measured in the lab, the launched DBS microwave
is a Gaussian beam with a beam radius of $a\approx3$ cm (i.e., the
radius where the beam power drops to $1/e^{2}$ of the maximum power).
Since the receiver CPS antenna is of the same design of the launching
DBS antenna, the scattered power that can be coupled to the receiver
antenna is determined by the convolution of the Gaussian beam profile
and the scattered wave power, which is proportional to $\exp\left(-2d^{2}/a^{2}\right)$.
Here, $d$ is the distance from the scattering center to the receive
beam axis. Due to this convolution effect, the received CPS signal
will also be affected by interference from multiple scattering centers.
These scattering points probably have different wavenumbers, local
mean densities, magnetic turbulence levels, and Doppler shifts. The
Doppler shifts can be used to separate signals from different locations.
Also, by employing the toroidal steering system, the $k_{\parallel}$
matching can be achieved for a single scattering location, hence the
signal from certain location can be enhanced over others.

Another effect that challenges the cross-polarization scattering measurement
is the mismatch between the polarization of the launched beam and
the plasma magnetic field pitch angle. The mismatch effect on the
CPS signal have been discussed in earlier studies.\citep{wang2018firststep,barada2016measurement,rhodes2014development}
Assuming an X-mode beam is launched, the mismatch of polarization
can result in a leaked O-mode component that propagates beyond the
X-mode cutoff layer, and give rise to an $O\rightarrow O$ backscattering
at the O-mode cutoff layer. The $O\rightarrow O$ backscattering leads
to a density fluctuation signal and hence contaminates the CPS signals.
Also, the pitch angle of the magnetic field is much larger in spherical
tokamaks than that in conventional tokamaks. Thus the mismatch effect
is more deleterious in STs, since the power of the unwanted/leaked
O-mode component is proportional to $\sin^{2}\delta$, where $\delta$
is the mismatch angle. For the combined DBS/CPS system in MAST-U,
this polarization matching issue has been addressed in CPS measurements
by using a remotely tunable polarizer to exclude the contamination
of the leaked mm-wave component.

\section{Summary and future work}

In this study, the ray-tracing analysis of the cross-polarization
scattering processes has been performed for the combined DBS/CPS diagnostics
on the MAST-U spherical tokamak, using the GENRAY 3D ray-tracing code.
Both the L-mode and H-mode cases have been considered, in which O-mode
CPS rays are emitted along the entire probing X-mode ray of DBS. The
matched wavenumbers of local magnetic turbulence are greater in the
H-mode than those in the L-mode, due to the large radial wavenumbers
in the H-mode case. The large radial wavenumbers probably lead to
smaller received CPS signal level. Some effects and challenges that
may affect the CPS signals are also discussed briefly. To improve
the detectable CPS signal level, a better match of the parallel wavenumber
$k_{\parallel}\approx0$ should be pursued by deploying a 2D launch/receive
antenna system. Also, the mismatch between the incident beam's polarization
and the magnetic pitch angle should always be avoided to prevent the
contamination due to the backscattered density fluctuations.

Furthermore, a full-wave simulation may be useful for a better understanding
of several effects that may complicate the interpretation of CPS signals,
e.g., the diffraction of the mm-wave beam, the size of scattering
volume, and interference of signals from different locations. Thus,
the 2D and 3D full-wave simulations will be pursued as the next step
for a synthetic CPS diagnostic.
\begin{acknowledgments}
The authors thank Dr.~N.~A.~Crocker and Dr.~Peng Shi for helpful
discussions. This material is based upon work supported by the U.S.
Department of Energy, Office of Science, Office of Fusion Energy Sciences
under Award~DE-SC0019005. 
\end{acknowledgments}

\section*{Data availability}

The data that support the findings of this study are available from
the corresponding author upon reasonable request.

\bibliographystyle{aipnum4-2}
\bibliography{ray_tracing_mastu}

\end{document}